\documentclass[reqno]{amsart}
\UseRawInputEncoding
\usepackage{graphics, graphicx}
\usepackage{amsfonts,amssymb,amsmath,amsbsy,amsthm,amscd}
\usepackage[english]{babel}
\usepackage{graphicx}
\newtheorem{theorem}{Theorem}[section]

\newtheorem{remark}{Remark}

\def\dfrac#1#2{\displaystyle{#1\over #2}}
\def\za{\alpha}
\def\zb{\beta}

\def\zo{\omega}

\def\bv{{\bf v}}
\def\bp{{\bf p}}

\def\Div{\mbox{div}\,}
\def\Rot{\mbox{curl}\,}

\def\bB{{\bf B}}

\def\bE{{\bf E}}

\begin{document}
\title[Exact thresholds in the dynamics of cold plasma with  collisions]{
Exact thresholds in the dynamics of cold plasma with electron-ion collisions}
\author{Olga S. Rozanova}
\address{ Mathematics and Mechanics Department, Lomonosov Moscow State University\\ Leninskie Gory,
Moscow, 119991,
Russian Federation\\
rozanova@mech.math.msu.su}

\author{Eugeniy V. Chizhonkov}

\author{Maria I. Delova}


\begin{abstract}
We consider a quasilinear system of hyperbolic equations that describes plane one-dimensional non-relativistic oscillations of electrons in a cold plasma with allowance for electron-ion collisions. Accounting for  collisions leads to the appearance of a term analogous to dry friction in a mechanical system, leading to a decrease in the total energy. We obtain a criterion for the existence of a global in time smooth solution to the Cauchy problem. It allows to accurately separate the initial data into two classes: one corresponds to a globally in time smooth solutions, and the other leads to a finite-time blowup. The influence of electron collision frequency  $ \nu $ on the solution is investigated. It is shown that there is a threshold value, after exceeding which the regime of damped oscillations is replaced by the regime of monotonic damping. The set of initial data corresponding to a globally in time smooth solution of the Cauchy problem expands with increasing $ \nu $, however, at an arbitrarily large value there are smooth initial data for which the solution forms a singularity in a finite time, and this time tends to zero as  $ \nu $ tends to infinity. The character of the emerging singularities  is illustrated by numerical examples.
 \end{abstract}


\maketitle

\section{Introduction}
Plasma is a highly nonlinear medium in which even relatively small initial collective displacements of particles lead to the excitation of oscillations and waves with a rather large amplitude. Their evolution can lead to the appearance of a singularity of the density of electrons~\cite{david72}.
This effect is called breaking of oscillations. As shown in \cite{ZM},
if the electron density goes to infinity in the Eulerian description of the motion, then the  electronic trajectories intersect in the Lagrangian description.

As a rule, the study of oscillations begins with a situation where electron-ion collisions are neglected.
 A criterion for the breaking of plane one-dimensional oscillations in the non-relativistic case  was obtained first in \cite{RozCh}. It confirmed the hypothesis formulated earlier \cite{Ch11}, \cite{Ch_book}.
When using Lagrangian variables, earlier attempts were also made to obtain a similar condition at the physical level of rigor, however, the reasoning was accompanied by technical errors~\cite{david72}.
The modern physical formulation of the problem  in the Lagrangian framework can be found in~\cite{M2013}.

In this paper, the results of ~\cite{RozCh} are extended to the case of taking into account electron-ion collisions.
It should be noted that the impact  of collisions on plasma oscillations has already been analyzed by other authors (see \cite{infeld},\cite{verma}).
However, in these studies, plasma resistance was taken into account simultaneously with viscosity.
For the first time, the effect of electron collisions together with the relativistic factor on the breaking of plane plasma oscillations was traced in \cite{FrCh}. However,  the analysis was carried out only by asymptotic and numerical methods, without exact mathematical formulations. In \cite{brodin},\cite{sahu} the effect of electronic collisions was analyzed in the nonrelativistic case, but only at a qualitative level.

The article has the following structure. First, a detailed statement of the Cauchy problem is given in Eulerian variables.
Then, the results on the formation of singularities at various values of the parameter $\nu$ characterizing the frequency of collisions are presented. It should be noted that the mathematical study covers all cases of non-negative values of $\nu$, although only small positive values have physical meaning. The following are numerical experiments illustrating the theoretical results. The initial conditions are selected as reasonable as possible from the point of view of full-scale physical experiments. For completeness, physical justification of the magnitude of the frequency of electron-ion collisions is given. In conclusion, the results of the studies are systematized.

\section{Formulation of the problem}

The system of hydrodynamics of a "cold" plasma, including hydrodynamic equations together with vectorial Maxwell's equations, has the form (see, e.g.~\cite{S71}, \cite{GR75}, \cite{SR12})
\begin{equation}
\label{base1}
\begin{array}{l}
\dfrac{\partial n }{\partial t} + \Div(n \bv)=0\,,\quad
\dfrac{\partial \bp }{\partial t} + \left( \bv \cdot \nabla \right) \bp
= e \, \left( \bE + \dfrac{1}{c} \left[\bv \times  \bB\right]\right) - \nu_e \bp,\vspace{0.5em}\\
\gamma = \sqrt{ 1+ \dfrac{|\bp|^2}{m^2c^2} }\,,\quad
\bv = \dfrac{\bp}{m \gamma}\,,\vspace{0.5em}\\
\dfrac1{c} \frac{\partial \bE }{\partial t} = - \dfrac{4 \pi}{c} e n \bv
 + {\Rot}\, \bB\,,\quad
\dfrac1{c} \frac{\partial \bB }{\partial t}  =
 - {\Rot}\, \bE\,, \quad \Div \bB=0\,,
\end{array}
\end{equation}
where
$e, m$ is the charge and mass of the electron (here the charge of the electron has a negative sign, $e < 0$),
$c$  is the speed of light,
$ n, \bp, \bv$ are density, momentum and speed of electrons,
$\gamma$ is the Lorentz factor,
$ \bE, \bB $ are vectors of electric and magnetic fields.

The left hand side of the equation for  momentum contains the term ${\bf g}=- \nu_e \bp $ describing electron-ion collisions.
The inclusion of this effect can be interpreted as the force of friction between particles,  see, for example,~\cite{ABR78} for the non-relativistic case. To describe collisions  they  often use the representation
$
{\bf g}=- \nu_{\za\zb} \left( \bv_{\za} - \bv_{\zb} \right),
$
where $\nu_{\za\zb}$ is the effective collision frequency of charged particles of  sort $\za$ with particles sort $\zb$, when $\za \neq \zb$.
For fixed ions ($ \bv_{\zb} = 0 $) the formula is simpler.

A detailed description of the formulas for plasma transport coefficients is presented in~\cite{b-20}.

In order to analyze plane one-dimensional non-relativistic plasma oscillations, the basic equations ~\eqref {base1} can be significantly simplified.

We denote the independent variables in the Cartesian coordinate system as
$ (x, y, z) $, and assume that the vector functions ${\bp}, {\bv}, {\bE}$ do not depend on
 $y$ and $z$. In addition, we formally set $\gamma \equiv 1,$ which  means neglecting the effects of relativism.

Then from~(\ref{base1}) we get
\begin{equation}
\begin{array}{c}
\dfrac{\partial n }{\partial t} +
\dfrac{\partial }{\partial x}
\left(n\, v_x \right)
=0,\quad
\dfrac{\partial E_x }{\partial t} = - 4 \,\pi \,e \,n\, v_x\,,\vspace{1 ex}\\
\dfrac{\partial p_x }{\partial t} + v_x \dfrac{\partial p_x}{\partial x}= e \,E_x - \nu_{e}p_x, \quad
{v_x} = \dfrac{p_x}{m}.
\end{array}
\label{3gl2}
\end{equation}
Then, we introduce dimensionless quantities
$$
\rho = k_p x, \quad \theta = \omega_p t, \quad
V = \dfrac{v_x}{c}, \quad
P = \dfrac{p_x}{m\,c}, \quad
E = -\,\dfrac{e\,E_x}{m\,c\,\omega_p}, \quad
N = \dfrac{n}{n_0}, \quad \nu = \dfrac{\nu_{e}}{\omega_p},
$$
 where $\omega_p = \left(4 \pi e^2n_0/m\right)^{1/2}$ is the plasma frequency, $n_0$ is the value of unperturbed electron density, $k_p = \omega_p /c$.
In the new variables we have $P=V$, and~\eqref{3gl2} takes the form
\begin{equation}
\dfrac{\partial N }{\partial \theta} +
\dfrac{\partial }{\partial \rho}
\left(N\, V \right)
=0,\quad
\dfrac{\partial V }{\partial \theta} + E +
V \dfrac{\partial V}{\partial \rho}  + \nu V = 0, \quad
\dfrac{\partial E }{\partial \theta} = N\, V\,.
\label{3gl3}
\end{equation}
From the first and last equations~\eqref{3gl3} we have
$$
\dfrac{\partial }{\partial \theta}
\left[ N +
\dfrac{\partial }{\partial \rho} E \right] = 0.
$$
This relation is valid both in the absence of plasma oscillations ($ N \equiv 1, \, E \equiv 0 $), and in their presence.
Therefore, we have a simpler expression for electron density
\begin{equation}
 N = 1 -
\dfrac{\partial  E }{\partial \rho}.
\label{gauss}
\end{equation}

Using it, we arrive at the equations describing
plane one-dimensional non-relativistic plasma oscillations taking into account collisions
\begin{equation}\label{q1}
\dfrac{\partial V }{\partial t}  + V \dfrac{\partial V}{\partial x}+ E+ \nu V = 0,\qquad
\dfrac{\partial E }{\partial t} + V \dfrac{\partial E}{\partial x} - V
=0.
\end{equation}
  Here, to go to the standard notation, we use
 $\rho = x\in \mathbb R$, $\theta = t\in \mathbb R_+$.  The functions $ V (x, t) $, $ E (x, t) $ have the meaning of velocity of electrons  and electric field strength, respectively, $ \nu $ is a fairly small non-negative constant characterizing the frequency of electron collisions.

   For the equations \eqref{q1} we consider the Cauchy problem
  \begin{equation}\label{qq1}
     (V(x,0),  E(x,0))= ( V_0(x), E_0(x)).
 \end{equation}
 In order to ensure the local-in-time existence of a classical ($ C^1 $) solution to the problem \eqref {q1}, \eqref {qq1}, it is necessary to require that the initial data belong to $ C^1 (\mathbb R) $. However, since our methods require an investigation of an extended system, we will require a greater smoothness, $ C^2 (\mathbb R) $.
 The aim of the research is to find conditions on the initial data under which the solution  loses smoothness over a finite time. This is important because, by virtue of the Gauss theorem, after breaking of oscillations, the mathematical model of cold plasma becomes inapplicable~\cite{Ch_book}.

  Problem \eqref {q1}, \eqref {qq1} for $ \nu = 0 $ was studied in~\cite{RozCh}, where a criterion for the loss of smoothness was found in terms of the initial data.

   In this paper, we solve a similar problem for $ \nu> 0 $ and investigate the influence of the coefficient $ \nu $ on the change in the region of the initial data for which the solution can be extended to the entire half axis $ t> 0 $.

  \section{A criterion for singularity formation}

  The system \eqref{q1} is hyperbolic, therefore the formation of singularities of its solution means that  either the solution itself or its derivatives go to infinity in a finite time \cite {Alinhac}. Below we show that in the case of the system \eqref{q1} the components of the solution remain bounded and tend to zero at infinity, but the derivatives can go to infinity.

  Let us denote $q=V_x$, $s=E_x$, $q_0(x) = q(x,0)$, $s_0(x)= s(x,0)$. Since the electron density has the form $ n = 1 - E_x$ (see \eqref{gauss}), then  only
$s_0<1$ has a physical sense.

Along the characteristics of  \eqref{q1}, the quantities $ V, E $ obey
  \begin{equation}\label{EV}
  \dot V= -E-\nu V, \quad \dot E = V,
 \end{equation}
and  $q,s$ obey
   \begin{equation}\label{sq}
  \dot q= -s-q^2-\nu q, \quad \dot s = q -q s.
 \end{equation}

An analysis of the phase plane made by standard methods shows that, depending on the value of the parameter $ \nu $, the following cases can be distinguished.

 \begin{itemize}
 \item 1. $\nu=0$: one equilibrium point $ (0,0) $, a center;
 \item 2. $0<\nu<2$: one equilibrium point $(0,0)$, a stable focus;
 \item 3. $\nu=2$:  two equilibria, $(0,0)$, a degenerate stable node, è $(1,-1)$, a saddle-node;
 \item 4. $\nu>2$:  three equilibria, $(0,0)$, a  stable node,  $(1,-\frac12(\nu-\sqrt{\nu^2-4})$, a saddle,  $(1,-\frac12(\nu+\sqrt{\nu^2-4})$, an  unstable node.
 \end{itemize}

Case 1 was considered in \cite{RozCh}. From the point of view of physical applications, the most interesting is the case 2 of  small  $ \nu $.

 The quadratic nonlinear system \eqref {sq} can be directly integrated. One possible way to do this is to reduce it to the Riccati matrix equation and further linearize \cite {Radon}. The solution is as follows:
   \begin{equation*}\label{s}
  s(t) = -\dot q(t)-q^2(t)-\nu q(t), \quad q(t)=\frac{G(t)}{F(t)},
 \end{equation*}
 where the form of the functions $ G $ and $ F $, $ F (0) \ne 0 $ depends on the case under consideration. The value of $ q $ (and $ s $ with it) goes to infinity in a finite time along the characteristic, starting from the point $ x_0 \in \mathbb R $ if and only if there exists $t_*>0$, such that $ F (t_*, x_0) = 0 $. The time when the derivatives of solution go to infinity is
  \begin{equation*}\label{Tc}
  T_{br}=\inf\limits_{t_*>0, \,x_0 \in  \mathbb R}  F(t_*, x_0)=0.
 \end{equation*}

In cases 1 and 2
  \begin{equation}\label{F2}
  F(t)= 1-s_0+ (A_{11} \,\sin \omega t + s_0 \,\cos \omega t )\,   e^{-\frac12 \nu t} ,\,
  \end{equation}
  \begin{equation*}\label{G2}
  G(t)= (- A_{12} \,\sin \omega t + q_0 \,\cos \omega t )\,   e^{-\frac12 \nu t},
  \end{equation*}
  \begin{equation*}\label{A}
    A_{11}=\frac{\nu s_0+2 q_0}{\sqrt{4-\nu^2}},\quad  A_{12}=\frac{\nu q_0+2 s_0}{\sqrt{4-\nu^2}} ,\quad \omega= \frac12\sqrt{4-\nu^2}.
 \end{equation*}
 In case 3
 \begin{equation}\label{F3}
  F(t)= 1-s_0+ (s_0+(s_0+q_0) t)\,   e^{- t} ,\,
  G(t)= -(s_0+(s_0+q_0) (t-1))\,   e^{- t},
  \end{equation}
In case 4
   \begin{equation}\label{F4}
  F(t)= 1-s_0+ (A_{21} \,\sinh \omega_1 t + s_0 \,\cosh \omega_1 t )\,   e^{-\frac12 \nu t} ,\,
 \end{equation}
  \begin{equation*}\label{G4}
  G(t)= (- A_{22} \,\sinh \omega_1 t + q_0 \,\cosh \omega_1 t )\,   e^{-\frac12 \nu t},
  \end{equation*}
  \begin{equation*}\label{s}
    A_{21}=\frac{\nu s_0+2 q_0}{\sqrt{\nu^2-4}},\quad A_{22}=\frac{\nu q_0+2 s_0}{\sqrt{\nu^2-4}},\quad\omega_1= \frac12\sqrt{\nu^2-4}.
 \end{equation*}
 In all these cases  $F(0)=1$, $G(0)=q_0$.

In terms of the phase portrait of  \eqref{sq}, we are looking for a necessary and sufficient condition for the boundedness of the phase curve. The phase curve itself can be found implicitly, however, the analysis of this expression is difficult.

Below we denote by $ \Phi_\nu (s_0, q_0) = 0 $ an implicitly defined phase curve in the plane $ (s_0, q_0) $ lying on the boundary of open region I and closed region II, where region I corresponds to points such that the phase curve starting from them tends to the origin in infinite time, and region II corresponds to points such that the phase curve starting from them tends to infinity during a finite time. Points lying on $ \Phi_\nu (s_0, q_0) = 0 $ itself also belong to closed region II.

If the initial data \eqref{qq1} is such that all points $ (s_0 (x_0), q_0 (x_0)) $, $ x_0 \in \mathbb R $, belong to region I, then the derivatives of the solution remain bounded and the solution is globally smooth by time. If there is at least one point $ x_0 \in \mathbb R $ such that $ (s_0 (x_0), q_0 (x_0)) $ falls into region II, then within a finite time (which can be found)  derivatives become infinite, that is, the solution loses smoothness.

 We also give the solution of system \eqref {EV}.

In cases 1 and 2 it is
 \begin{equation*}\label{E12}
 E(t)= (\bar A \,\sin \omega t + E_0 \,\cos \omega t )\,   e^{-\frac12 \nu t} ,\quad \bar A=\frac{\nu E_0+2 V_0}{\sqrt{4-\nu^2}},
 \end{equation*}
  \begin{equation}\label{V12}
 V(t)= (\bar B \,\sin \omega t + V_0 \,\cos \omega t )\,   e^{-\frac12 \nu t} ,\quad \bar B=-\frac{2 E_0+\nu V_0}{\sqrt{4-\nu^2}};
 \end{equation}
in cases 3
 \begin{equation}\label{EV3}
 E(t)= (E_0+(V_0+E_0)t) e^{-t}, \quad V(t) = (V_0-(V_0+E_0)t) e^{-t};
 \end{equation}
 in cases 4
 \begin{equation*}\label{E4}
 E(t)= (\bar A_1 \,\sinh \omega_1 t + E_0 \,\cosh \omega_1 t )\,   e^{-\frac12 \nu t} ,\quad \bar A_1=\frac{\nu E_0+2 V_0}{\sqrt{\nu^2-4}},
 \end{equation*}
  \begin{equation}\label{V4}
 V(t)= (\bar B_1 \,\sinh \omega_1 t + V_0 \,\cosh \omega_1 t )\,   e^{-\frac12 \nu t} ,\quad\bar B_1=-\frac{2 E_0+\nu V_0}{\sqrt{\nu^2-4}}.
 \end{equation}

We see that in all cases, provided that the smoothness is maintained, the components of the solution tend to zero as $ t \to \infty $ .

 \subsection{ Case $0<\nu<2$}


 We find the necessary and sufficient conditions to vanish  for $ F (t) $, given as \eqref{F2}. Note that the amplitude of oscillation of $ F (t) $ decreases with increasing $ t $, therefore, if it does not vanish during the first period of the function $ A \, \sin \omega t + s_0 \, \cos \omega t $, starting from $ t = 0 $ , it will never vanish.
The point at which $ F (t) $ has an extremum can be found in the standard way, on the interval $ (-\frac{\pi}{\sqrt {4-\nu^2}},\frac{\pi}{\sqrt {4-\nu^2}} )$ it is
\begin{equation*}\label{T_ext}
 T_{ext}=  T_1= {\frac {2}{\sqrt {4-\nu^2}}}\,\arctan \left({\frac {{q_0}\sqrt {4-\nu^2}}{\nu\,q_0+2
\, s_0}} \right).
 \end{equation*}
The value of the second derivative of $ F (t) $ at this point is
\begin{equation*}\label{minmax}
 -{\rm sign}(\nu q_0+2 s_0)\, \sqrt{q_0^2+\nu\, q_0\,s_0+s_0^2}
\,\,{\exp}\left(-\frac{\nu}{2}\,T_1 \right).
 \end{equation*}
 Therefore, $ T_1 $ is a minimum point only for $\nu q_0+2 s_0<0$, and for $ \nu q_0 + 2 s_0> 0 $ the minimum point is the following extremum point,  $T_2=T_1+\frac{2\pi}{\sqrt {4-\nu^2}}$. If $T_1<0$, we consider the next minimum point on the positive semi-axis, $T_3=T_1+\frac{2\pi}{\sqrt {4-\nu^2}}$. We must find the values of $ (s_0, q_0) $ for which $ F (t) $ at the minimum point equals to zero.

 Thus, for $ \nu q_0 + 2 s_0 <0 $, $q_0<0$, the curve separating in the plane $ (s_0, q_0) $ the initial data for which the derivatives of the solution always remain smooth, and for which the solution forms a singularity during a finite time, is
 \begin{equation}\label{f1}
F(T_1)=  1-s_0 -  \sqrt{q_0^2+\nu\, q_0\,s_0+s_0^2}
\,\,{\exp}(-\frac{\nu}{2}\,T_1 )=0.
 \end{equation}
 For $ \nu q_0 + 2 s_0> 0 $ such a curve is
 \begin{equation}\label{f2}
F(T_2)=  1-s_0 -  \sqrt{q_0^2+\nu\, q_0\,s_0+s_0^2}
\,\,{\exp}(-\frac{\nu}{2}\,T_2 )=0,
 \end{equation}
 for $ \nu q_0 + 2 s_0 <0 $, $q_0>0$, it is
\begin{equation}\label{f3}
F(T_3)=  1-s_0 -  \sqrt{q_0^2+\nu\, q_0\,s_0+s_0^2}
\,\,{\exp}(-\frac{\nu}{2}\,T_3 )=0.
 \end{equation}
 Note that $ \lim \limits_{s_0 \to - \frac {\nu} {2} q_0-0} T_1 = \lim \limits_{s_0 \to - \frac {\nu} {2} q_0 + 0} T_2 = \frac {\pi} {\sqrt {4- \nu ^ 2}} $, $q_0<0$, and $ \lim \limits_{s_0 \to - \frac {\nu} {2} q_0-0} T_2 = \lim \limits_{s_0 \to - \frac {\nu} {2} q_0 + 0} T_3 = \frac {\pi} {\sqrt {4- \nu ^ 2}} $, $q_0<0$, so the curve  on the plane $ (s_0, q_0) $, composed of $ F (T_1) = 0 $, $ F (T_2) = 0 $ and $ F (T_3) = 0 $, is continuous at the point of intersection with $ \nu q_0 + 2 s_0 = 0 $. It is easy to see that it is also smooth at this point. Thus, the curve $ \Phi_\nu (s_0, q_0) = 0 $ is obtained; it is presented in Figure 1 on the left.

\begin{center}
\begin{figure}[h]
\begin{minipage}{0.495\columnwidth}
\centerline{
\includegraphics[width=0.9\columnwidth]{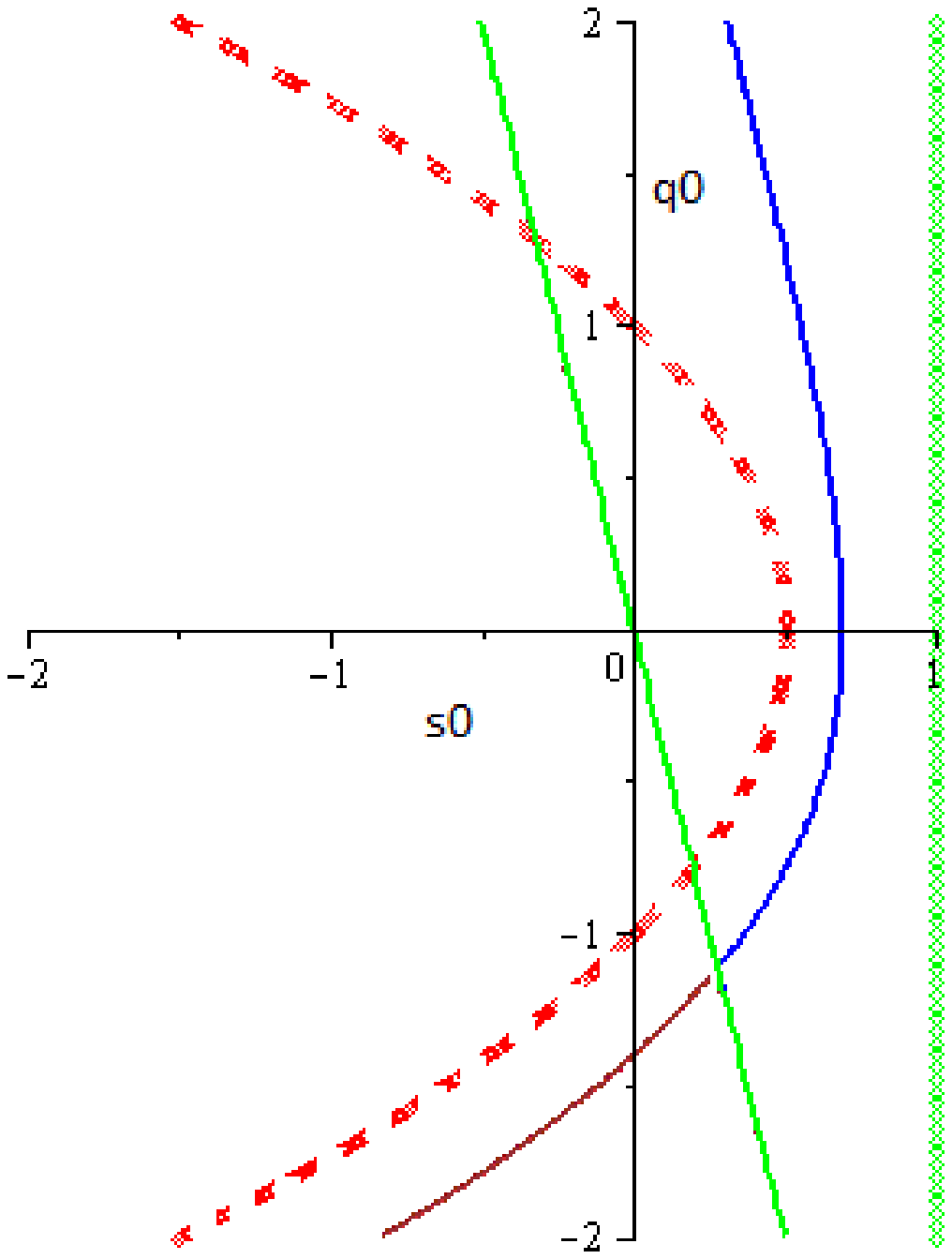}}
\end{minipage}
\begin{minipage}{0.495\columnwidth}
\centerline{
\includegraphics[width=0.9\columnwidth]{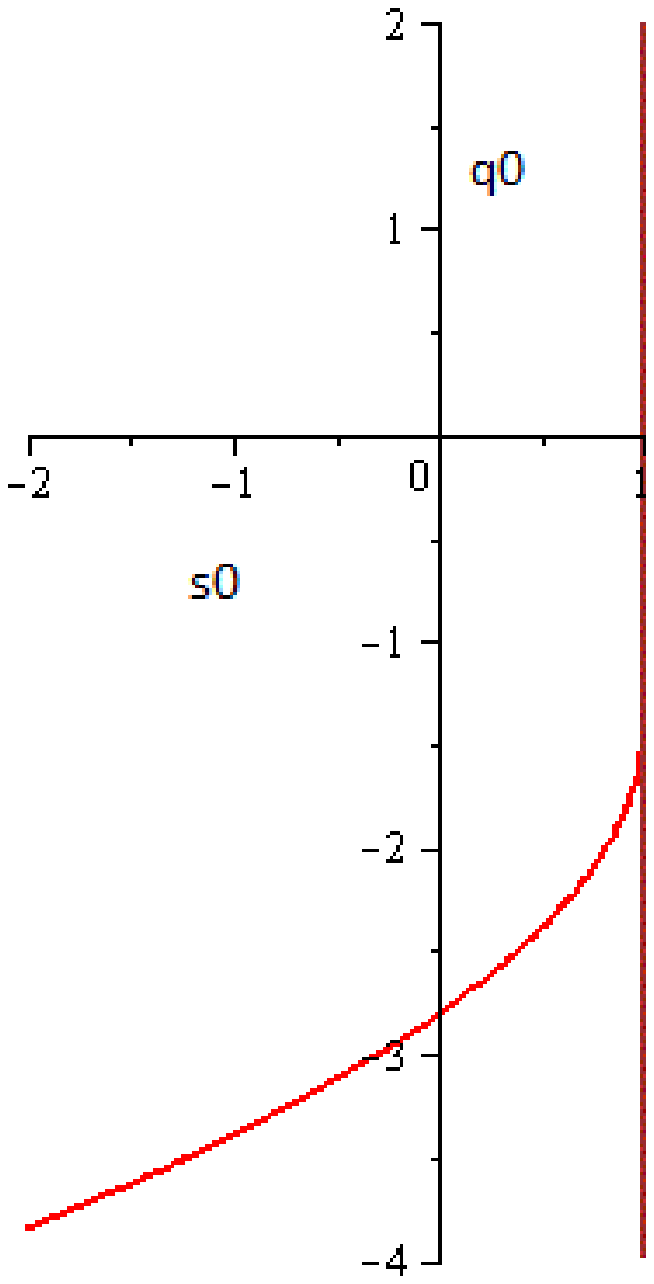}}
\end{minipage}
\caption{Left: The graph of $ \Phi_ \nu (s_0, q_0) = 0 $ and the line $ \nu q_0 + 2 s_0 = 0 $ at $ \nu = 0.5 $ in comparison with the graph of $ \Phi_\nu (s_0, q_0) = 0 $ for $ \nu = 0 $ (dotted line). Right: Graph of the function $ \Phi_\nu (s_0, q_0) = 0 $ at $ \nu = 2.1 $ with respect to the line $ s_0 = 1 $.}
\end{figure}
\end{center}

 Thus, we obtain the theorem

\bigskip
\begin{theorem}
Let $ 0 <\nu <2 $. If initial data \eqref {qq1} is such that for all $ x_0 \in \mathbb R $ the point $ (s_0 (x_0), q_0 (x_0)) $ lies in that part of the half-plane $ s_0 <1 $, separated by the curve $ \Phi_\nu (s_0, q_0) = 0 $ (given by \eqref {f1}, \eqref {f2}), where the coordinate origin falls, then the classical solution to problem \eqref{q1}, \eqref{qq1} exists for all $ t> 0 $. Otherwise, for some finite time, the derivatives of the solution of problem \eqref {q1}, \eqref {qq1} become infinite.
\end{theorem}

\begin{remark}
In the limit  $ \nu \to 0 $ we can get the equation of the curve $ \Phi_\nu = 0 $ as $ 1-s_0 - \sqrt{q_0^2 + s_0^2} = 0
$ or $ q_0^2 + 2s_0-1 = 0 $, which  corresponds to the result \cite{RozCh}.
\end{remark}

\begin{remark}
 A study of the region on the plane $ (q_0, s_0) $, corresponding to a globally smooth solution, shows that it expands with increasing $ \nu $.
\end{remark}

 \subsection{ Case $\nu=2$}

\begin{theorem}
Let $ \nu = 2 $. If  initial data \eqref {qq1} is such that for all $ x_0 \in \mathbb R $ the point $ (s_0 (x_0), q_0 (x_0)) $ lies in that part of the half-plane $ s_0 <1 $, separated by the curve $ \Phi_\nu (s_0, q_0) = 0 $ given by \eqref{phi2}, where the coordinate origin is, then the classical solution to problem \eqref {q1}, \eqref {qq1} exists for all $ t> 0 $. Otherwise, for some finite time, the derivatives of the solution  become infinite.
\end{theorem}
\bigskip
\proof
 The extremum of the function $ F (t) $ defined as \eqref {F3} is at the point
 $$T_{ext}=\ln  \left( \frac { s_0+q_0}{ s_0-1} \right),
  $$
  it is always a minimum. This value is positive for $ q_0 <-1 $. Therefore, the equation of the curve $ \Phi_\nu (s_0, q_0)=0 $ has the form
  $
  F(T_{ext})=0,
  $
  or
  \begin{equation}\label{phi2}
   (s_0+q_0)\ln  \left( \frac { s_0+q_0}{ s_0-1} \right)- q_0=0,\,q_0<-1.
   \end{equation}
  This curve coincides with one of the phase curves of  \eqref {sq}, leaving the equilibrium  $ (1, -1) $,
   $$ (s_0+q_0)\ln  \left( \frac { s_0+q_0}{ s_0-1} \right)=-s_0+C (q_0+s_0),$$ for $C=1$. If $C<1$, then this phase curve comes to the equilibrium  $ (0,0) $ and it is bounded. The theorem is proved.

\bigskip
\begin{remark}
 The case $ \nu = 2 $ is transitional from the regime of damped oscillations of a smooth solution at $ \nu <2 $, to the non-oscillatory behavior of smooth solutions at $ \nu> 2 $.
\end{remark}

 \subsection{ Case $\nu>2$}

\begin{theorem}
 Let $ \nu> 2 $. If initial data \eqref {qq1} is such that for all $ x_0 \in \mathbb R $ the point $ (s_0 (x_0), q_0 (x_0)) $ lies in that part of the half-plane $ s_0 <1 $, separated by the curve $ \Phi_\nu (s_0, q_0) = 0, $ given as \eqref{F4T_ext}, $ q_0 <0, $ $ q_0 <- \frac {2} {\nu + \sqrt {\nu^2-4}} s_0 $, where the origin is, then the classical solution to problem \eqref{q1}, \eqref {qq1} exists for all $ t> 0 $. Otherwise, for some finite time, the derivatives of the solution  become infinite.
\end{theorem}
 \bigskip
\proof
 We study the possibility of vanishing of the function $ F (t) $ given by  \eqref {F4}. It reaches a minimum at a single extremum point
 $$T_{ext}=\frac{1}{2 \omega_1}\ln  \left( \frac { s_0^2+s_0 q_0 z + q_0^2 (\frac{z\nu}{2} -1)}{ s_0^2 +\nu s_0 q_0+q_0^2} \right),\,z=\nu+\sqrt{\nu^2-4}.
  $$
  This expression is positive provided $q_0(s_0+\frac{z}{2}q_0)>0$. Graph of the function
  \begin{equation}\label{F4T_ext}F(T_{exp})=0
  \end{equation}
  lies in the domain $q_0<0, q_0<-\frac{2}{z} s_0$, the curve is outgoing from the unstable equilibrium
$(1,-\frac12(\nu+\sqrt{\nu^2-4})$. The region of the phase plane corresponding to a globally smooth solution lies above it, see Fig.1 on the right. The theorem is proved.

   \subsection{Estimate of the density of a smooth solution}

   Since the electron density is expressed  as $\, n = 1 - s $, it is easy to calculate what  minimum value $ n $ can achieve  if the solution maintains global smoothness. To do this, let us find the maximal possible value of $ s $. It is easy to see that for $ \nu \ge 2 $ there are no additional restrictions, i.e. $ n> 0 $. For $ 0 \le \nu <2 $, one must find the maximum possible value of $ s $ on the curve $ \Phi_\nu = 0. $
    Since
    $\dfrac{d s_0(q_0)}{d q_0}=-\frac{(\Phi_\nu)_{q_0}}{(\Phi_\nu)_{s_0}}$, then at the point of extremum $(\Phi_\nu)_{q_0}=0$. It is easy to calculate that this happens when $ q_0 = 0 $. Since under this condition $s_0=\frac{1}{1+\exp(-\frac{\nu \pi}{\sqrt{4-\nu^2}})}$, then for a globally smooth solution the estimate holds:
    \begin{equation*}\label{n}
n>\frac{\exp(-\frac{\nu \pi}{\sqrt{4-\nu^2}})}{1+\exp(-\frac{\nu \pi}{\sqrt{4-\nu^2}})}\to 0, \quad \nu\to 2.
\end{equation*}
Results of this kind are very rare, and therefore are of particular value for the construction of approximate methods for solving hyperbolic systems.

 \section{Numerical experiments}

The main "working tool" during the experiments was the second-order McCormack scheme ~\cite{Ch20} on a grid uniform in both independent directions with constant steps $ \tau $ and $ h $ in time and space, respectively.

 To control calculations, we used the  algorithm in Lagrangian variables, described in~\cite{FrCh}, adapted to the non-relativistic case.

 \subsection{Data selection}

Let the velocity at $t=0$ be
\begin{equation}
V_0(x) = 0,
\label{3gl6}
\end{equation}
and we assume that the initial an electric field is
\begin{equation}\label{datal2}
E_0(x) = k\, x \,\exp\left\{-
\dfrac{x^2}{\sigma}\right\}, \quad k = \left(\dfrac{a_*}{\rho_*}\right)^2, \; \sigma=\frac{\rho_*^2}{2}.
\end{equation}
The form of~(\ref{datal2}) was chosen for reasons that such oscillations can be excited in a rarefied plasma by a laser pulse with a frequency of $ \zo_l $ ($ \zo_l \gg \zo_p $) when it is focused in a line, which can be achieved by using a cylindrical lens~\cite{Shep13}.

The parameters $ \rho_* $ and $ a_* $ characterize the scale of the localization region and the maximum value
$ E_{\max} = a_*^2 / (\rho_* 2 \sqrt {{\rm e}}) \approx 0.3 a_*^ 2 / \rho_* $ of the electric field, respectively.
The expression~(\ref{datal2}) models the displacement of electrons at the initial time relative to the point $ x = 0 $ in different directions, which subsequently leads to their oscillations, due to Coulomb interactions with stationary ions.

The following are physical arguments for estimating the value of the coefficient $ \nu $ of the electron collision frequency.

The following relation is valid~\cite{GFCA}:
\begin{equation}
   a_*^2 = a_0^2 \tau_*\sqrt{\pi/2}  \exp\left(-\tau_*^2/8\right),
\label{3gl9}
\end{equation}
   here $a_0 = eE_{0L}/(m \zo_l c)$ is the normalized amplitude of the laser field, $E_{0L}$ is the physical amplitude of the laser field. Under conditions of optimal excitation of the wake wave ($ \tau_* = 2 $), when its amplitude is maximum, ~(\ref {3gl9}) takes the form $a_*^2 = a_0^2 \sqrt{2\pi/{\rm e}}\approx 1.52 a_0^2$.

In a fully ionized plasma, the dimensionless frequency of electron-ion collisions is given by~\cite{SR12}
\begin{equation}
 \nu= Z\dfrac{\sqrt{8}}{3}\eta^{3/2} \ln \Lambda,
\label{3gl11}
\end{equation}
where $Z$  is the ion charge number, $\ln \Lambda$ is the Coulomb logarithm, and
the parameter $ \eta $ is the ratio of the electron interaction energy $e^2n^{1/3}_{0}$ to the electron kinetic energy $T_e$.

Suppose a laser pulse propagates in a fully ionized underdense plasma with an ion charge $ Z = 5 $, whose electrons have a density  $n_{0} = 10^{18} \,\mbox{\rm sm}^{-3} $
and temperature $T_e = 50$ eV.  Its wavelength is $ \lambda_l = 1.24 $ $\rm \mu m$
( frequency is $\omega_l = 1.5 \times 10^{15} \,\mbox{\rm s}^{-1}$), duration is $\tau = 36$ fs and dimensionless electric field amplitude is $a_0=2.5$.
If the laser pulse is focused by a cylindrical lens in a line with a transverse dimension $L_x = 24$ $\rm \mu m$, then mainly plane one-dimensional oscillations are excited in the wake wave behind the pulse,   $a_* = 3.1, \, \rho_* =4.5$  (it corresponds $k \approx 0.4746$). These parameters are close to those used in the computations.

For the given plasma parameters, the dimensionless collision frequency can be found from~(\ref{3gl11}), it is $\nu = 0.5 \times 10^{-2}$.
If we consider the propagation of a laser pulse with the above parameters in a plasma with the same density, but with a temperature $T_e = 20$ eV,
then calculations by formula~(\ref {3gl11}) give for the dimensionless collision frequency the value $\nu = 1.8 \times 10^{-2}$.
Thus, the cold plasma approximation allows one to study the effect of electron collisions on plasma oscillations, although for any physically attainable plasma parameters this effect is quite small ($ \nu \ll 1$).

In the numerical simulation of plasma oscillations, the computational domain must be limited. To this end
note that at large distances from the point $ x = 0 $, due to the initial condition~\eqref {datal2}, the deviations of the initial data from zero are very small. Indeed, the characteristic trajectory along which the values of $ E, V $ change, modulo exponentially decaying, obeys the equation $ \dot x (t) = V (t) $, where $ V (t) $ is known (see \eqref {V12}, \eqref {EV3}, \eqref {V4}). Therefore, $ | x (t) -x (0) | $ is uniformly bounded and can be estimated. Therefore,
if
$\lim\limits_{|x| \to \infty}E_0(x) =\lim\limits_{|x| \to \infty} V_0(x)=0,$
so that we can assume that the initial data outside a certain interval $ [- d_0, d_0] $ is negligible, then
\begin{equation*}
\lim\limits_{|x| \to \infty}E(x,t) =\lim\limits_{|x| \to \infty} V(x,t)=0.
\label{3gl10}
\end{equation*}
Then there exists an interval $ [- d, d] $, outside which the solution can be considered negligibly small to the same extent. We consider this interval as the computational domain. At the ends of $ [- d, d] $, artificial boundary conditions should be specified.
Section 3.6 in~\cite {Ch_book} is devoted to the discussion of their construction, here we restrict ourselves to "cutting off" an infinite region using homogeneous boundary conditions of the first kind:
$$
 V(\pm d,t) =  E(\pm d,t) = 0.
$$
Of course, the parameter $ d $ should be chosen large enough.
For  \eqref {3gl6}, \eqref {datal2}, we just put $ d = 4.5 \rho_* $. In this case, we have $\, \exp (- d^2 / \sigma) \approx 2.5768 \cdot 10^{- 18} $. This means that, in double-precision calculations, the value of jump  of the initial function $ E_0 $ at the points $ x = \pm d $ is commensurate with machine accuracy, i.e. with the usual error of data rounding. In other words, in the numerical simulation of oscillations, the effect of cutting off the initial conditions will not be noticeable at all, which fully corresponds to the concept of an ``artificial boundary''.

This approach is practically very convenient and therefore the most frequently used, but its main drawback is an excessive increase in the computational domain.
The observed effect of oscillation breaking, as a rule, is realized in the vicinity of the origin of the coordinate $ x $ at a distance of less than $ 0.1 \, \rho_* $, therefore, more than $ 90 $ percent of the calculations are a kind of ``payment'' for using the ``rough'' boundary conditions. In this case, the optimization of computations is not the goal of the work; therefore, the simplest version of the boundary conditions is quite suitable for illustrating the properties of the solutions under discussion.

In addition, we take into account the invariance property of the derivatives of the solution with respect to the change of coordinate (see section 3.2.2 in~\cite{Ch_book}). Thus, when the parameter $ k $ in the initial condition~(\ref {datal2}) is changing,  it is  possible to confine oneself only to the value $ \rho_* = 1 $ without losing the content of numerical experiments. For example, the set of values $ a_* = 3.1, \, \rho_* = 4.5 $ in the considered non-relativistic case is quite acceptable to replace with the set  $a_* = 0.69, \, \rho_* =1.0$.

Fig.2 illustrates the qualitative behavior of the solution for various values of $ \nu $. It depicts a dependency
electron density versus time for $ x = 0 $ for a fixed value of $ k = 0.4761 $. In accordance with the above theoretical results
at any value of $ \nu $ from the set of $ 0.0, \, 0.2, \, 2.2 $, there is no breaking, however at $ \nu = 0.0 $ we observe a $ 2 \pi- $ periodic solution, at $ \nu = 0.2 $ it is a damped oscillatory solution, for $ \nu = 2.2 $ it is a monotonically damped (non-oscillating) solution.

\medskip
\begin{center}
\begin{figure}[h!]
\centerline{
\includegraphics[width=0.7\columnwidth]{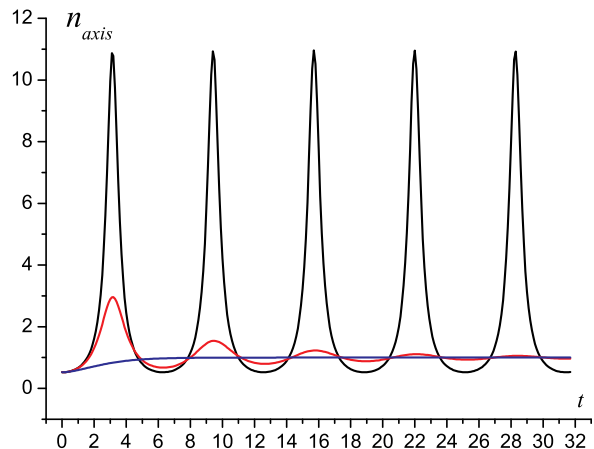}}
\caption{
 Dependence of electron density on time at $ x = 0 $ at various intensities of collisions: $ \nu = 0 $ (black), $ \nu = 0.2 $ (red), $ \nu = 2.2 $ (blue).}
\end{figure}
\end{center}

 \subsection{Computions for $0<\nu<2$}

When choosing the initial conditions~(\ref{3gl6}), we have $ q_0 = 0 $ for all $ x_0 $. It should be noted that with an increase in the amplitude $ k $, the point that first falls on the curve separating  the region  of preservation of smoothness from the region of formation of singularities on the plane $ (s_0, q_0) $ is $ (s_0 (0), 0 ) $. Moreover, it is easy to calculate the critical value of $ k$,
 \begin{equation*}\label{k_kr}
k_{cr}=\frac{1}{1+\exp(-\frac{\nu \pi}{\sqrt{4-\nu^2}})}
\end{equation*}
and the time  of the singularity formation,
 \begin{equation*}\label{k_kr}
T_{br}=\frac{2 \pi}{\sqrt{4-\nu^2}}.
\end{equation*}

 Fig.3  presents the dependence of the breaking time  on the amplitude $ k $ at various collision frequencies. It is easy to see that in the absence of collisions ($ \nu = 0 $), the critical value of $ k $ is equal to $ k_{cr} = 1/2 $, the corresponding  $ T_{br} = \pi $. At lower values of $ k $, the breaking does not occur. On the other hand, if we approach arbitrarily close to  $ k = 1 $, it cannot accelerate the breaking process compared to the time $ T_{br} = \pi / 2 $. Recall that the value $ k = 1 $ corresponds to $ s_0 = 1 $, which in turn generates the minimum density value $ n = 0 $, i.e. lack of electrons. The above means that if the electrons are completely removed from any subdomain, then the consequence of such a situation will be a breaking over time $ T_{br} = \pi / 2 $ (in the absence of collisions).

If the collisions are taken into account, then for an arbitrary initial amplitude of oscillations $ k <1 $, the breaking time increases in comparison with the case $ \nu = 0 $,
which follows from the higher position of the curve for $ \nu> 0 $. Moreover, the breaking time $ T_{br} $ decreases  with an increase of $ k_{cr} $ (the latter is strictly greater than $ 1/2 $ for any $ \nu $). Therefore, in Fig.3, the domain of function  $ T_{br} (k) $ starts from $ 1/2 $.

\begin{center}
\begin{figure}[h!]
\centerline{
\includegraphics[width=0.7\columnwidth]{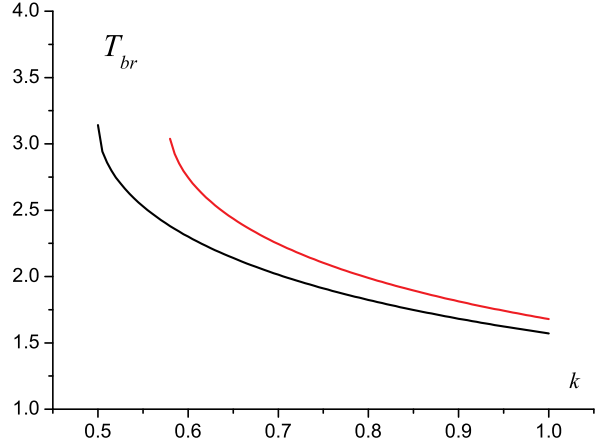}}
\caption{
 The dependence of the breaking time on  $ k $ at different collision frequencies: $ \nu = 0.0 $ (black), $ \nu = 0.2 $ (red).}
\end{figure}
\end{center}
Fig.4 shows a typical damped oscillatory process for small values of the parameter $ \nu $. In particular, for $ \nu = 0.2 $, for the oscillation amplitude $ k = k_{cr} $, we consider the transformation of a smooth initial function $ E_0 (x) $ into a function $ E_{br} (x) $, in which the derivative is singular. In full accordance with the analytical derivations, this happens at the time $ t = T_{br} $. We note that, by virtue of taking into account electronic collisions, the range of values of the final function is smaller than that of the initial one, although the oscillatory process is observed quite distinctly.

\begin{center}
\begin{figure}[h!]
\centerline{
\includegraphics[width=0.7\columnwidth]{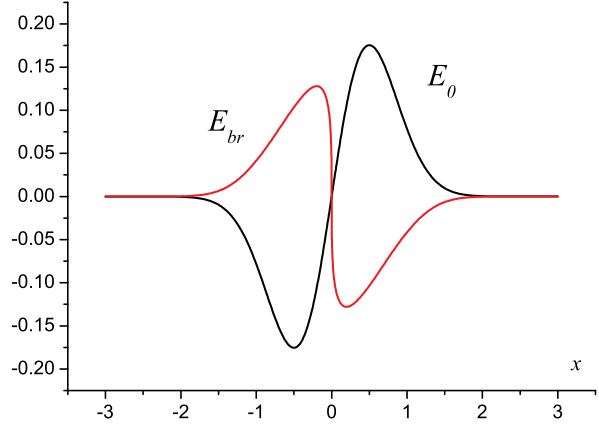}}
\caption{Transformation of a smooth initial electric field (black) into a stepwise function at $ x = 0 $ at the time of breaking for $ \nu = 0.2 $ (red).}
\end{figure}
\end{center}

\bigskip

 \subsection{Computations for $\nu >2$}

We show that for large $ \nu $ there are initial data that lead to the formation of singularities. However, now this cannot be achieved by leaving the initial speed unperturbed. We choose the initial data of the following form:
\begin{equation*}\label{datag2}
E_0(x) = k_1 x \exp\left\{-
\dfrac{x^2}{\sigma}\right\},\quad V_{0}(x) =  -k_2 x \exp\left\{-
\dfrac{x^2}{\sigma}\right\}, \quad k_1, k_2>0.
\end{equation*}
The analytical expression \eqref {F4T_ext} for a phase curve that bounds the region of initial data corresponding to a smooth solution is very cumbersome. Therefore, for simplicity, we choose $ k_1 = 0. $ In this case, calculations show that the threshold value $k_2 = k_{cr} $ is given by the formula
$$
k_{cr}=\frac{\sqrt{\nu^2-4}}{\exp(z_+m)-\exp(z_-m)},$$$$z_\pm=\frac{-\nu\pm\sqrt{\nu^2-4}}{2},\quad m=\frac{1}{\sqrt{\nu^2-4}}\,\ln \frac{\nu\sqrt{\nu^2-4}+\nu^2-2}{2}.
$$
The quantity $ k_{cr} $ tends to infinity with increasing $ \nu $. The time of formation of the infinite derivative  is equal to
$$T_{br}=m.$$ It monotonically decreases with $ \nu $ and tends to zero as $ \nu $ tends to infinity.

A clear illustration of the non-oscillatory breaking is shown in Fig.~5. If we choose a sufficiently large value of the collision frequency (in the fugure $ \nu = 2.5 $, which is practically devoid of physical meaning), then a numerical experiment allows us to observe the formation of singularity in the electric field function from the zero initial condition $ E_0 (x) $. The moment of breaking is $ t = T_{br}. $ This effect is achieved only due to the initial electron velocity at $ k_2 = k_{cr} $ and it is monotonous (non-oscillatory) in nature. The figure shows how the derivative of electric field at  $ x = 0 $   changes from zero (at $ t_0 = 0 $) to minus-infinity (at $ t= T_{cr} $) with conservation of sign, in contrast with the oscillatory process. The graphs are shown at time instants $ t_1 = T_{cr} / 4 $, $ t_2 = T_{cr} / 2 $, $ t_3 = T_{cr} $.

\begin{center}
\begin{figure}[h!]
\centerline{
\includegraphics[width=0.7\columnwidth]{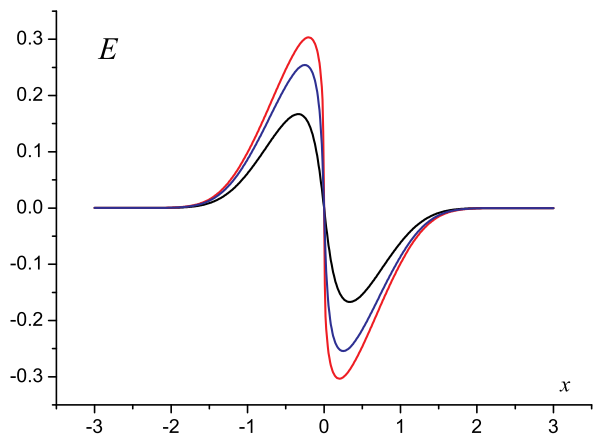}}
\caption{
 Non-oscilatory breaking for $ \nu = 2.5 $. Monotonic decrease of the derivative of the electric field  $ E (x, t) $ at $ x = 0 $ for time instants: $ t_1 = T_{br} / 4 $ (black), $ t_2 = T_{br} / 2 $ (blue), $ t_3 = T_{cr} $ (red).}
\end{figure}
\end{center}

\section{Conclusion}

We study the possibility of the existence of a smooth solution of a hyperbolic system  describing plane oscillations in an electron plasma with allowance for electron collisions. For the first time, a full analytical study of the model was carried out.
It is shown that with an increase in the frequency of electron collisions, the region of initial data corresponding to a globally smooth solution expands. However, for an arbitrarily large value of the collision frequency coefficient $ \nu $, it is impossible to obtain a smooth solution for any smooth initial data. For large values of $ \nu $, the solution either monotonically decays or loses smoothness very quickly. Moreover, there is a threshold value $ \nu = 2 $ at which the attenuation pattern changes: for $ 0 <\nu <2 $ oscillations occur, and for $ \nu> 2 $ the attenuation is monotonous.

The results can be useful in the design and justification of high-precision numerical algorithms, as well as for modeling
 oscillatory processes in a cold plasma, taking into account the effects of relativism, the influence of a magnetic field, viscosity and other physical factors.

\end{document}